\documentstyle[12pt,psfig]{article}
\begin{document}
\def \beq{\begin{equation}}
\def \eeq{\end{equation}}
\rightline{EFI-95-22}
\rightline{CERN-TH/95-145}
\rightline{ENSLAPP-A-523/95}
\rightline{ISN 95-07}
\rightline{ITKP 95-17}
\rightline{hep-ph/9506315}
\vspace{0.5in}
\centerline{\bf SPACINGS OF QUARKONIUM LEVELS}
\centerline{\bf WITH THE SAME PRINCIPAL QUANTUM NUMBER
\footnote{To be submitted to Phys.~Rev.~D.}}
\vspace{0.5in}
\centerline{\it Aaron K. Grant and Jonathan L. Rosner}
\centerline{\it Enrico Fermi Institute and Department of Physics}
\centerline{\it University of Chicago, Chicago, IL 60637}
\bigskip
\centerline{\it Andr\'e Martin}
\centerline{\it Theory Group, CERN, 1211-CH-Geneva 23, Switzerland}
\medskip
\centerline{\it and}
\medskip
\centerline{\it Laboratoire de Physique Th\'eorique ENSLAPP\footnote{\rm ~URA
14-36 du CNRS associ\'ee \`a l'\'Ecole Normale Sup\'erieure de Lyon et \`a
l'Universit\'e de Savoie}}
\centerline{\it Groupe d'Annecy, LAPP, B. P. 110}
\centerline{\it F-74941 Annecy-le-Vieux, France}
\bigskip
\centerline{\it Jean-Marc Richard}
\centerline{\it Institut de Sciences Nucl\'eaires--CNRS-IN2P3}
\centerline{\it Universit\'e Joseph Fourier}
\centerline{\it 53, avenue des Martyrs, 38026 Grenoble Cedex, France}
\medskip
\centerline{\it and}
\medskip
\centerline{\it Institut f\"ur Theoretische Kernphysik}
\centerline{\it Rheinische Friedrich-Wilhelms Universit\"at}
\centerline{\it Nu$\ss$allee 14-16, D-53115 Bonn, Germany}
\bigskip
\centerline{\it Joachim Stubbe}
\centerline{\it D\'epartement de Physique Th\'eorique, Universit\'e de
Gen\`eve}
\centerline{\it CH-1211 Gen\`eve 4, Switzerland}
\newpage

\centerline{\bf ABSTRACT}
\medskip
\begin{quote}
The spacings between bound-state levels of the Schr\"odinger equation with the
same principal quantum number $N$ but orbital angular momenta $\ell$ differing
by unity are found to be nearly equal for a wide range of power potentials
$V = \lambda r^\nu$, with $E_{N \ell} \approx F(\nu, N) - G(\nu,N) \ell$.
Semiclassical approximations are in accord with this behavior. The result is
applied to estimates of masses for quarkonium levels which have not yet been
observed, including the 2P $c \bar c$ states and the 1D $b \bar b$ states.
\end{quote}
\bigskip

\centerline{\bf I.  INTRODUCTION}
\bigskip

The properties of heavy quarkonium ($c \bar c$ and $b \bar b$) levels have
provided useful insights into the nature of the strong interactions.  At
short distances these interactions appear to be characterized by a Coulombic
potential associated with asymptotically free single-gluon exchange \cite{AF},
while at large distances the interquark force is consistent with a constant,
corresponding to a separation energy increasing linearly with distance
\cite{LD}. An effective power potential $V \sim r^\nu$, with $\nu$ close to
zero, provides a useful interpolation between these two regimes for $c \bar c$
and $b \bar b$ levels when the interquark separation ranges between about 0.1
and 1 fm \cite{PL}.

Charmonium ($c \bar c$) levels have been identified up to the fourth or fifth
S-wave, the first P-wave, and the first D-wave excitation.  (We are ignoring
fine structure and hyperfine structure for the moment.)  Six S-wave levels and
two P-wave levels have been found in the upsilon ($b \bar b$) family.  Many
additional levels are expected but have not yet been seen.  There are some
reasons of current interest for predicting their positions in a relatively
model-independent manner.  For example:

(1)  It has been suggested \cite{Wise,Close} that, depending on its exact mass,
the second charmonium P-wave level (which we shall denote as $\chi_c'(2P)$)
could play a role in the hadronic production of the $\psi'(2S)$ state.

(2)  The first D-wave level of the $b \bar b$ system, which we shall call
$\Upsilon(1D)$, may be accessible to experiments of improved sensitivity at the
CLEO detector, both in direct production of the $^3D_1$ states via the $e^+
e^-$ channel and through electromagnetic transitions from the $\Upsilon(3S)$
state \cite{KR}.

Explicit calculations in nonrelativistic quark models of the masses of the
$\chi_c'(2P)$ and $\Upsilon(1D)$ states have tended to have very small spreads
\cite{KR,preds}.  Any interquark potential which reproduces the known
quarkonium levels is well-enough determined to leave little room for variation
in predictions of these levels.  However, in the course of examining these
predictions within the context of power potentials $V = \lambda r^\nu$,
with $\lambda \nu > 0$, we were struck by a curious feature:  The spacings
between levels with the same {\it principal quantum number} but orbital
angular momenta differing by unity are nearly identical for a wide range of
values of $\nu$.

The principal quantum number $N$ is that which labels nonrelativistic Coulomb
energy levels through the Balmer formula $E_{N \ell} \propto -1/N^2$.  It is
related to the radial quantum number $n_r$ and the orbital angular momentum
$\ell$ via $N = n_r + \ell ~(n_r = 1,2,3,\ldots)$, where $n_r - 1 \equiv n$
corresponds to the number of nodes of the radial wave function between $r = 0$
and $r = \infty$.  Our result amounts to a formula for levels $E_{N \ell}$ {\it
linear in $\ell$} for a given $N$:
\begin{equation} \label{eqn:en}
E_{N \ell} = F(\nu, N) - G(\nu, N) \ell
\end{equation}
with $G(\nu,N) > 0$ for $\nu > -1$ in accord with a much more general result
which specifies the order of levels for fixed $N$ as a function of $\ell$
\cite{GM}.  This general result is that $G$ is positive for any potential whose
Laplacian is positive everywhere outside the origin.  We investigate this
general case first in Section II, turning to the special case of power
potentials in Sec.~III.  We apply the results to specific quarkonium cases in
Sec.~IV, while Sec.~V summarizes.  Some proofs of identities are given in three
appendices.
\bigskip

\centerline{\bf II.  POTENTIALS WITH POSITIVE LAPLACIAN}
\bigskip

We consider the Schr\"odinger equation in appropriate dimensionless units
($\hbar = 2 \mu = 1$, where $\mu$ is the reduced mass), for a spherically
symmetric potential $V(r)$:
\beq \label{eqn:schr}
[- \Delta + V(r)] \Psi_{N \ell}({\bf r}) = E_{N \ell} \Psi_{N \ell}({\bf r})
{}~~.
\eeq

Potentials with positive Laplacian $\Delta V > 0$ play a special role in
quarkonium physics. First, all potentials used to fit the $b \bar b$ and $c
\bar c$ spectrum have positive Laplacian (except for spurious pathologies when
one expands the potential in powers of the strong coupling constant at a given
scale).  It is in fact very natural to take potentials with positive Laplacian
because this property is a kind of expression of asymptotic freedom:  If we say
that the force between a quark and an antiquark is $F = - \alpha(r)/r^2$,
asymptotic freedom requires that $\alpha(r)$ be an increasing function of $r$.
Writing $F = -dV/dr$, saying that $\alpha(r)$ increases is equivalent to
\beq
\frac{d}{dr} r^2 \frac{dV}{dr} > 0~~~,
\eeq
which means that $V$ has a positive Laplacian.  If this is the case we have
\cite{GM}, for a given multiplet,
\beq
\label{eqn:multiplet-ordering}
E_{N,\ell-1} > E_{N,\ell} > E_{N,\ell+1}
\eeq
for $N \ge 3,~\ell \ge 1$.

Very soon after the discovery of this property, Baumgartner \cite{Baum}
proved the following theorem:

Define
\beq
R_{N,\ell} \equiv \frac{E_{N,\ell-1} - E_{N,\ell}}{E_{N,\ell} - E_{N,\ell+1}}
{}~~.
\eeq
Then
\beq \label{eqn:rineq}
R_{N,\ell} \ge \frac{\ell}{\ell+1} \frac{N^2 - (\ell+1)^2}{N^2 - \ell^2}~~~,
\eeq
if $V = -(1/r) + \lambda v$, with $\Delta v \ge 0,~\lambda$ sufficiently
small.  Notice that the right-hand side of (\ref{eqn:rineq}) is always less
than unity.  So (\ref{eqn:rineq}) holds for {\em perturbations} of a
Coulomb potential.  It is not known if this inequality also holds outside the
perturbative regime.  Our belief is that it does hold, but the proof must be
rather complex.  The perturbative proof is based on the equation
$$
\ell \alpha^2_{\ell-1} (E_{N,\ell-1} - E_{N,\ell})
- (\ell + 1) \alpha^2_{\ell} (E_{N,\ell} - E_{N,\ell+1})
$$
\beq \label{eqn:Coul}
= \frac{2 \ell + 1}{2} \int_0^{\infty} \Delta V(r) u_{N,\ell}^2 dr~~~
\eeq
where $u_{N,\ell}$ is a pure Coulomb wave function, and
\beq
4 \alpha_{\ell}^2 \equiv (\ell+1)^{-2} - N^{-2}~~~.
\eeq
For  $\Delta V >0$, the inequality (\ref{eqn:rineq}) follows from
(\ref{eqn:Coul}) and (\ref{eqn:multiplet-ordering}). In fact, inequality
(\ref{eqn:rineq}) would also hold for a potential with purely negative
Laplacian, since both (\ref{eqn:Coul}) and (\ref{eqn:multiplet-ordering})
are reversed.
 The proof of
(\ref{eqn:Coul}) will be given in Appendix A.  What is important is that this
equation shows that Baumgartner's result \cite{Baum} cannot be improved,
because it is saturated by a potential $v$ whose Laplacian has its support
concentrated at the zeroes of $u_{n,\ell}$.  Such a potential is easy to
construct:  It is made of piecewise shifted Coulomb potentials.  The resulting
overall potential is not concave, but can be made so by a correction of order
$\lambda^2$ ($\lambda$ being the coefficient of $v$), for example by joining
pairs of Coulomb segments by linear pieces tangent to each.

Even if one believes that (\ref{eqn:rineq}) holds for non-perturbative
situations, it is a rather disappointing result for practical applications.
For instance, using the ``particle physics'' spectroscopic notation, we get
\beq
\frac{E_{3S} - E_{2P}}{E_{2P} - E_{1D}} > \frac{5}{16}~~~,
\eeq
while, as we shall see in the next section, this ratio can be very close
to unity.  On the other hand no improvement is possible.  Even the very
simple-minded potential $-(1/r) + \lambda r$ gives
\beq
\frac{E_{3S} - E_{2P}}{E_{2P} - E_{1D}} \to \frac{1}{2}
\eeq
for $\lambda \to 0$, as shown in Appendix A, and we have checked numerically
that for finite $\lambda$ this limit is approached smoothly.  Therefore, to
make progress, we should use a more restricted class of potential, the power
potentials, which are known to give excellent fits of the heavy quark-antiquark
spectra \cite{PL}.
\newpage

\centerline{\bf III.  THE CASE OF POWER POTENTIALS}
\bigskip

If $V(r) = \varepsilon(\nu)r^\nu$, where $\varepsilon(\nu)$ is the sign
function and $-1 < \nu < 2$, one finds that $R_{N,\ell}$ is always very close
to unity, to summarize briefly this section.  The evidence comes from
indications from semi-classical formulae valid for large quantum numbers,
analytical investigations in the neighborhood of $\nu = -1$ and $\nu = 2$, and
explicit numerical calculations of energy levels for small quantum numbers,
shown in Fig.~1.  For the purpose of this figure, in order to obtain a smooth
limit as $\nu \to 0$, we have represented the Schr\"odinger equation as
\beq \label{eqn:schrfig}
[- (1/2)\Delta + (r^\nu - 1)/\nu]
\Psi_{n_r \ell}({\bf r}) = E_{n_r \ell} \Psi_{n_r \ell}({\bf r})~~~,
\eeq
and levels are labeled by the radial quantum number $n_r = N - \ell$ and the
orbital angular momentum $\ell = (0,~1,~2,~3,~\ldots) = $(S, P, D, F,
$\ldots$).
\begin{figure}
\psfig{file=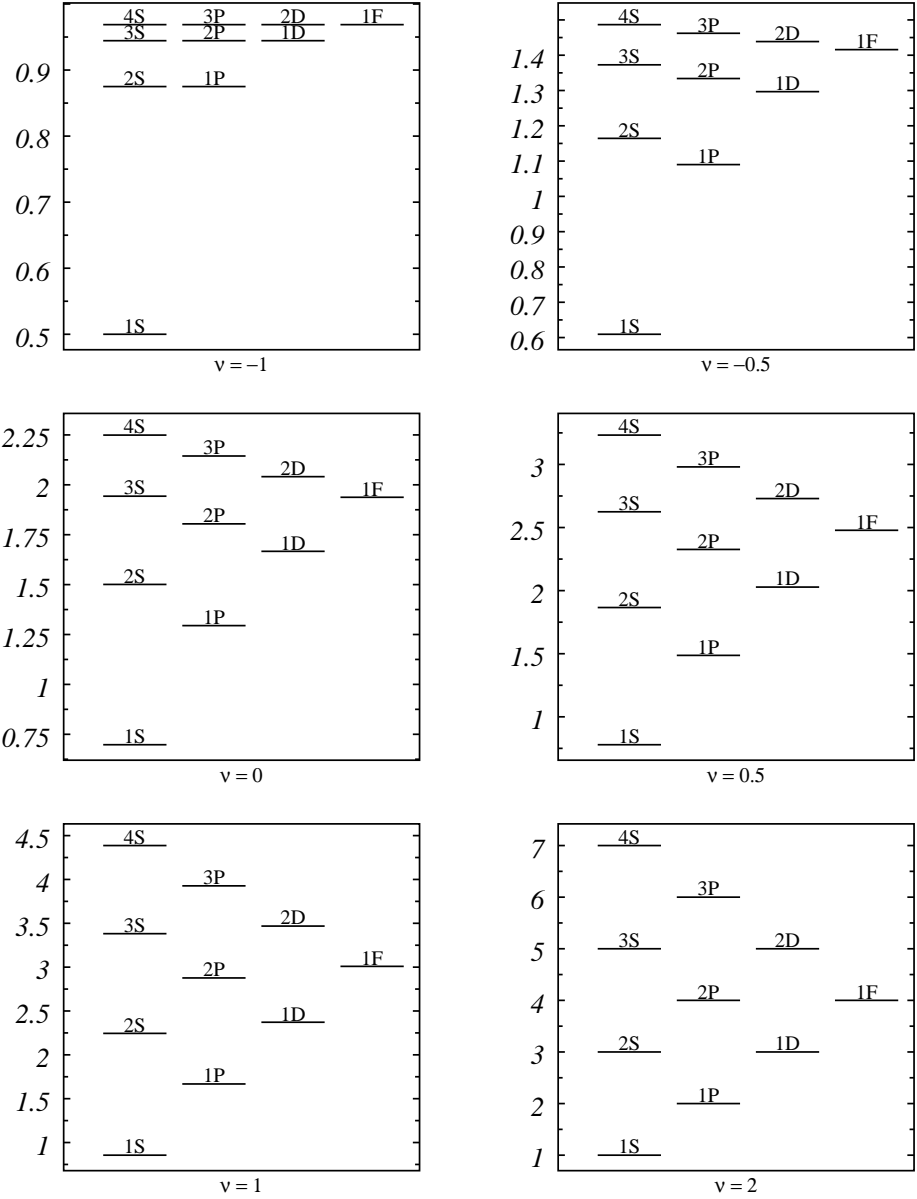,width=5.5in}
\caption{Energy levels in the Schr{\"o}dinger equation
  (\protect{\ref{eqn:schrfig}}) for several power potentials.  Levels
  are labeled as described in the text.}
\end{figure}

For power potentials, semi-classical formulae are expected to hold for $N \to
\infty$.  However there are several regimes depending on whether $\ell$ stays
small compared to $N$ or, on the contrary, $N - \ell$ stays small. In the
latter case two of us \cite{GR} have shown that $E_{N,\ell}$ depends
asymptotically only on the combination $N - \ell(1 - \frac{1}
{\sqrt{\nu + 2}})$. In the former case $N \to \infty,~\ell$ finite,
$E_{N,\ell}$ depends on $N - \frac{\ell}{2}$, the same combination as in the
case of the harmonic oscillator, as shown by one of us and C. Quigg \cite{QR}.

An exhaustive study of the various cases has been made by Feldman, Fulton, and
Devoto \cite{FFD}.  We shall not give in the text the refined formulae obtained
by these authors.  (A further refinement is given in Appendix B.)  We shall
content ourselves with the leading terms.  Here we express energies in terms of
the number of nodes $n = n_r - 1 = N - \ell - 1$ and the angular momentum
$\ell$.  We have, defining $E(N,\ell) \equiv \hat E(n,\ell)$, the
following cases:
\medskip

\leftline{\it i) $\ell$ large, $n$ finite, $-2 < \nu < \infty$:}
$$
\hat E(n,\ell) \sim \varepsilon(\nu) \left( \frac{ |\nu|}{2}
\right)^{-\frac{\nu}{\nu+2}} \left( 1 + \frac{\nu}{2} \right)
$$
\beq \label{eqn:largel}
\times \left[ \ell + \frac{1}{2} + \left(n + \frac{1}{2} \right)
\sqrt{\nu + 2} \right]^{\frac{2 \nu}{\nu + 2}}~~~,
\eeq

\leftline{\it ii~a) $n$ large, $\ell$ finite \cite{QR}, $0 < \nu < \infty$:}
\beq
\hat E(n,\ell) \sim \left[ 2 \sqrt{\pi} \frac{\Gamma(\frac{3}{2} +
\frac{1}{\nu})} {\Gamma( 1 + \frac{1}{\nu})} \left( n + \frac{\ell}{2} -
\frac{1}{4} \right) \right]^{\frac{2 \nu}{\nu+2}}~~~,
\eeq

\leftline{\it ii~b) $n$ large, $\ell$ finite \cite{QR}, $-2 < \nu < 0$:}
\beq
\hat E(n,\ell) \sim - \left\{ 2 |\nu| \sqrt{\pi} \frac{\Gamma(1 -
\frac{1}{\nu})} {\Gamma(- \frac{1}{2} - \frac{1}{\nu})} \left[ n - \frac{1}{2}
\left( \frac{1 + \nu - 2 \ell}{2 + \nu} \right) \right] \right \}
^{\frac{2 \nu}{\nu + 2}}~~~.
\eeq

These formulae exhibit very smooth behavior of the energies as functions of
$n$ and $\ell$ and hence $N$ and $\ell$.  Further support for the
smoothness in $\ell$ for fixed $n$ is given in Appendix B, because
$[\hat E(n,\ell)]^{\frac{\nu+2}{\nu}}$ is a ``Herglotz'' function of $(\ell
+ 1/2)^2$ for $\nu \ge 0$.

In the case {\it i)}, for $\ell \to \infty$, $n$ finite, one can write a
systematic expansion of the energies in inverse powers of $\ell + \frac{1}
{2}$ \cite{GMNP}.  The first terms are
$$
\hat E(n,\ell) = \left( \frac{|\nu|}{2} \right) ^{\frac{2}{\nu+2}}
\frac{2+\nu}{\nu} \left( \ell + \frac{1}{2} \right)^{\frac{2 \nu}{\nu + 2}}
$$
$$
\times \left\{1 + \frac{2 \nu}{\sqrt{\nu+2}} \frac{n + \frac{1}{2}}{\ell
+ \frac{1}{2}} + \frac{\nu(\nu-2)}{12(\nu+2)(\ell+\frac{1}{2})^2}
\left[ (11- \nu)(n + \frac{1}{2})^2 + \frac{\nu+1}{12} \right] \right.
$$
\beq \label{eqn:ehatexpn}
\left . + {\cal O} \left[ \frac{\nu - 2}{(\ell+ \frac{1}{2})^3} \right]
\right\}~~~.
\eeq
Notice that this expression becomes exact for $\nu = 2$.  In Appendix B we show
that (\ref{eqn:ehatexpn}) can be used to improve the Feldman--Fulton--Devoto
\cite{FFD} asymptotic expression for large $\ell$ in a rather impressive way.
Eq.~(\ref{eqn:ehatexpn}) yields the behavior of $R_{N,\ell}$ for $\ell \to
\infty,~N-\ell$ finite:
$$
R_{N,\ell} \simeq 1 + \frac{(\nu-2)^2}{6(2\ell+1)}\frac{4 - 3 \sqrt{\nu+2}
- \nu}{(\nu+2)(\sqrt{\nu+2}+2)}
$$
\beq \label{eqn:Rexpn}
+ {\cal O} \frac{ (\nu-2)^2}{(2\ell+1)^2}~~~.
\eeq
This expression shows that $R_{N,\ell}$ is indeed very close to unity
for $-1 \le \nu \le 2$, as long as $\ell \ge 2$.

Other limiting cases which can be treated analytically are $\nu \to -1$ and
$\nu \to 2$, for arbitrary $n$ and $\ell$.

In the case $\nu \to -1$ it is sufficient to look at a perturbation of the
Coulomb potential by a potential $\log(r)/r$, since
\beq
\lim_{\nu \to -1} \frac{r^\nu - r^{-1}}{\nu+1} = \frac{\log(r)}{r}
{}~~.
\eeq
Calculations presented in Appendix C give
\beq \label{eqn:Rcoul}
R_{N,\ell} = \frac{N+\ell+1}{N+\ell}
\eeq
for $\nu \to -1$.

In the case $\nu \to 2$, one should, similarly, take a perturbation
$V = r^2 \log r$.  Then one gets the surprise that the terms of order $\nu-2$
vanish and
\beq
R_{N,\ell} \simeq 1 + {\cal O}(\nu-2)^2~~~.
\eeq

At this point, we see that it may be advantageous to replace
Eq.~(\ref{eqn:Rexpn}) by
\beq
R_{N,\ell} \simeq 1 + \frac{(\nu-2)^2}{6(N+\ell)}\frac{4 - 3 \sqrt{\nu+2}
- \nu}{(\nu+2)(\sqrt{\nu+2}+2)}~~~,
\eeq
which preserves the large-$\ell$ asymptotic behavior and agrees with
(\ref{eqn:Rcoul}) for $\nu \to -1$ and (\ref{eqn:Rexpn}) for $\nu \to 2$.

All this fits perfectly into the numerical calculations presented in Figs.~2
and 3 for $(N=3,~\ell=1)$, $(N=4,~\ell=1)$, and $(N=4,~\ell=2)$.  It is
striking that the asymptotic formula (\ref{eqn:Rexpn}) agrees quite well with
the case in which $n$ is smallest and $\ell$ largest, i.e., $N=4,~\ell=2$.  In
particular the asymptotic formula reproduces the fact that $R-1$ is negative
for $\sim 0.1 < \nu < 2$.

\begin{figure}
\psfig{file=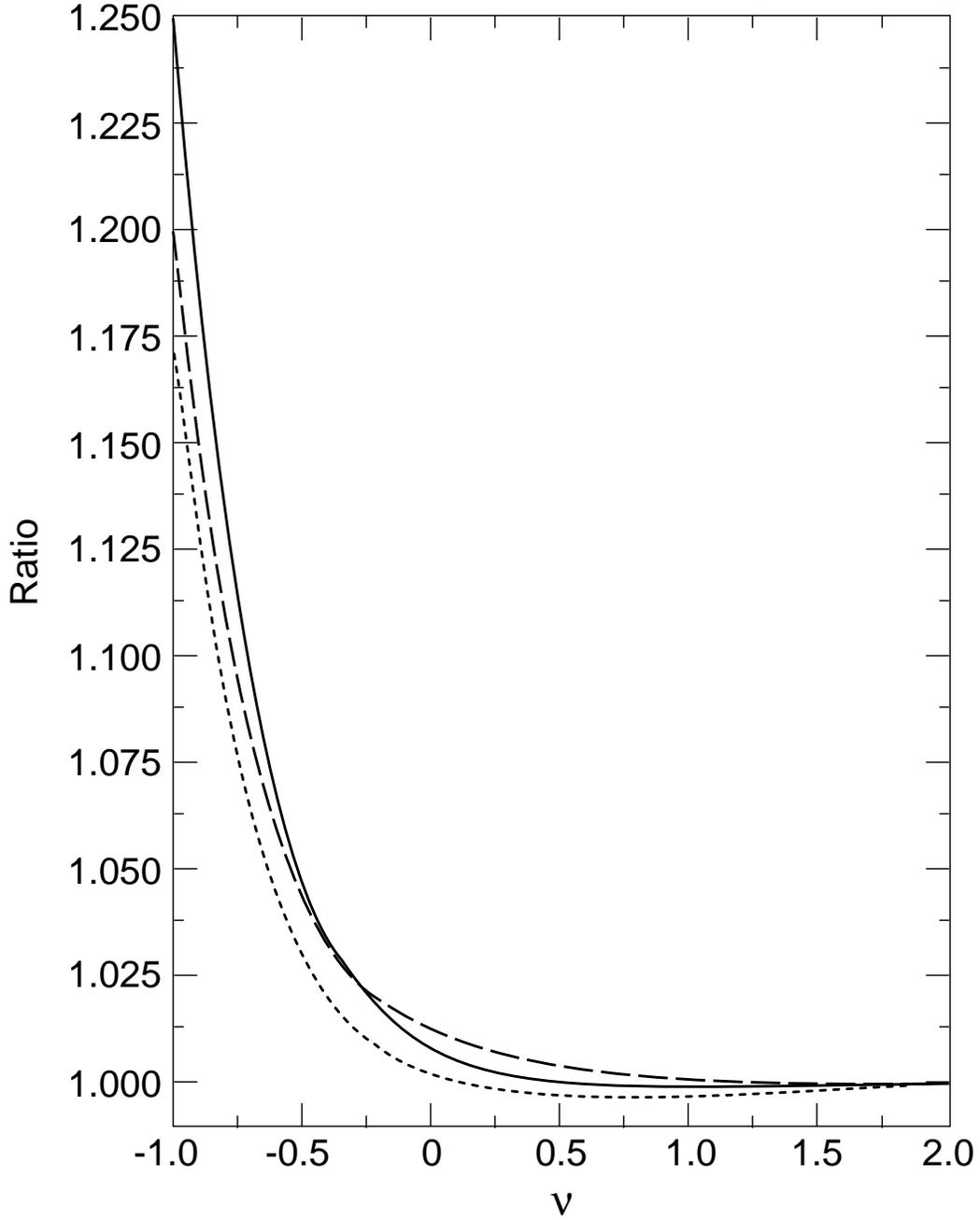,width=5.5in}
\caption{Ratios of spacings of levels with the same principal quantum number in
power-law potentials $V(r) \sim r^\nu$ as a function of $\nu$. Solid curve:
$[E(3S) - E(2P)]/[E(2P) - E(1D)]$; Dashed: $[E(4S) - E(3P)]/[E(3P) - E(2D)]$;
Dotted: $[E(3P) - E(2D)]/[E(2D) - E(1F)]$. Here levels are labeled by $n_r$(S,
P, D, F, $\ldots$), where $n_r = N - \ell$ is the radial quantum number.}
\end{figure}

\begin{figure}
\psfig{file=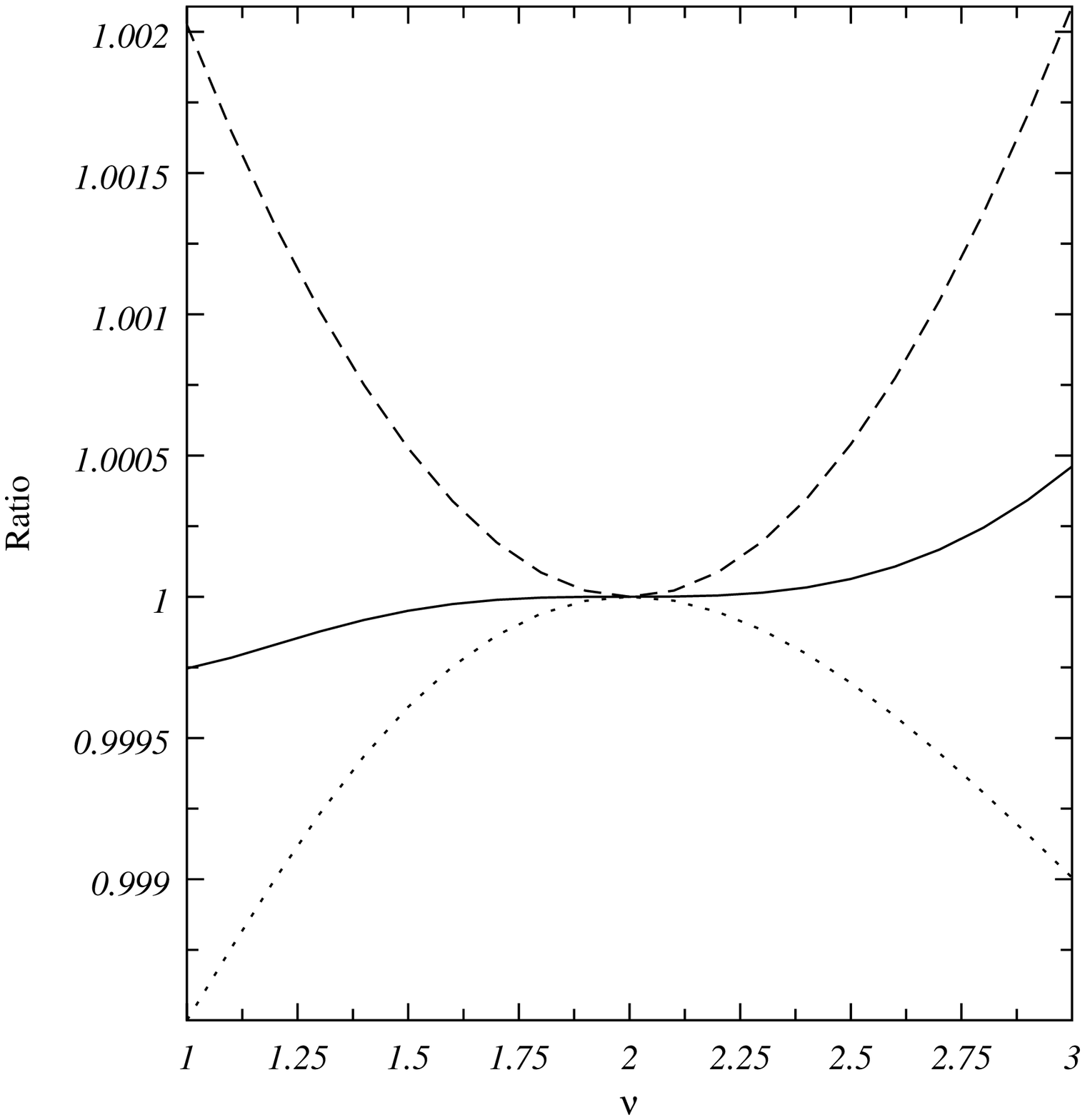,width=5.5in}
\caption{Ratios of spacings of levels with the same principal quantum number in
power-law potentials $V(r) \sim r^\nu$ near $\nu = 2$. Solid line: $[E(3S) -
E(2P)]/[E(2P) - E(1D)]$; dashed line: $[E(4S) - E(3P)]/[E(3P) - E(2D)]$; dotted
line: $[E(3P) - E(2D)]/[E(2D) - E(1F)]$. Levels are labeled as in Fig.~2.}
\end{figure}

For completeness let us indicate that for $\nu \to -2$, $R_{N,\ell}$ tends to
$+ \infty$. Notice first that, by scaling, $R_{N,\ell}$ is independent of the
strength of the power potential.  We can always take $V = - g r^\nu$ with
$(\ell+\frac{1}{2})^2 > g > (\ell-\frac{1}{2})^2$.  Then $E_{N,\ell}$ and
$E_{N,\ell+1}$ approach zero as $\nu \to -2$, while $E_{N,\ell-1}$ goes to
$-\infty$.  This is because, for a state of angular momentum $\ell$, the limit
Hamiltonian is, for $\nu \to -2$,
\beq
H_{\rm lim} = - \frac{d^2}{dr^2} - \frac{1}{4r^2} + \frac{(\ell+\frac{1}{2})^2
- g}{r^2}~~~.
\eeq
The operator $-(d^2/dr^2) - 1/(4r^2)$ is known to be positive, i.e., its
expectation value for any reduced wave function vanishing at the origin is
positive. Hence, if $g < (\ell + \frac{1}{2})^2$ the eigenvalues of $H_{\rm
lim}$ are positive. However, for $0 > \nu > -2$, the Hamiltonian has negative
eigenvalues (in fact infinitely many!) and therefore the lowest eigenvalue has
to go to zero for $\nu \to -2$.  On the other hand for $g > (\ell +
\frac{1}{2})^2$ it is known that $H_{\rm lim}$ is not lower-bounded, as can be
seen by taking its expectation value with, for instance, a trial function
$u = 0$ for $r \le R_m$, $(r - R_m)^{\frac{1}{2} + \epsilon}(R_M - r)$ for
$R_m < r < R_M$, 0 for $r \ge R_M$, and letting $R_m$ and $\epsilon$ go to
zero.  This not only proves that the ground state energy goes to $- \infty$,
but {\em all} radial excitation energies do so as well.  If $E_n$ is a finite
limiting value of the energy level of the $n$th radial excitation, there is a
sequence of energies and wave functions (defined by integration of the
Schr{\"o}dinger equation from infinity)
which approaches the limit energy and wave function.  Let $r_k$ be the first
nonzero limit of a node for $\nu \to -2$.  Then taking $r_{k-1} < R_m < R_M
< r_k$, the energy corresponding to the interval $(R_m,R_M)$ with Dirichlet
boundary conditions is higher than the one corresponding to $(r_{k-1},r_k)$
and goes to $- \infty$, which is what we wanted to prove.  If all nodes
approach zero (which is in fact the case!) the proof obviously works, taking
$R_m > r_n$.

The opposite extreme case is $\nu \to \infty$, i.e., the square well.  Then the
solutions of the Schr\"odinger equation are Bessel functions which can be
approximated near their turning point by Airy functions.  One gets, for
instance,
\beq
\lim_{\ell \to \infty} R_{\ell+2,\ell} = \frac{|a_3| - |a_2|}
{|a_2| - |a_1|} = 0.818709 \ldots~~~,
\eeq
where the $a_i$ are successive zeros of the Airy function.
\bigskip

\centerline{\bf IV.  APPLICATIONS TO QUARKONIUM LEVELS}
\bigskip

Realistic quarkonium potentials appear to have a power-law behavior $V \sim
r^\nu$, \cite{PL}, with $\nu$ not far from zero.  For $-0.1 < \nu < 0.1$,
$R_{3,1}$ is $1.007 \pm 0.002$.  Consequently, one can anticipate that energy
levels with the same principal quantum number $N$ will depend nearly linearly
on the orbital angular momentum $\ell$.  We apply this result to two examples,
one in charmonium and one in the upsilon family.
\bigskip

\leftline{\bf A.  Charmonium $2P$ states}
\bigskip

The $1P$ levels of charmonium were identified nearly 20 years ago in electric
dipole transitions from the $\psi'(2S)$ \cite{oldP}.  Recently they have been
studied with great precision in proton-antiproton formation experiments
\cite{E760}.  Since they lie below threshold for decay to a pair of charmed
mesons, they are quite narrow, facilitating their observation.  The $^3P_1$ and
$^3P_2$ levels, in particular, have substantial branching ratios to $\gamma +
J/\psi$.

Recent interest has focused on the possibility that one or more charmonium $2P$
states may be narrow enough to have a substantial branching ratio to $\gamma +
\psi'(2S)$ \cite{Wise,Close}.  This suggestion is motivated by a production
rate for $\psi'(2S)$ in high-energy proton-antiproton collisions \cite{CDFpsi}
which appears too high to be explained by conventional $QCD$ mechanisms.  It
then becomes of some interest to predict exactly where the $2P$ levels should
lie. If the $^3P_2$ level lies sufficiently low, its decay to $D \overline{D}$,
though kinematically allowed, will be suppressed by a large $\ell = 2$
centrifugal barrier, so that its branching ratio to $\gamma + \psi'(2S)$ could
be non-negligible.  The $^3P_1$ level cannot decay to $D \overline{D}$; it will
be narrow if it lies below the $D \overline{D}^*$ threshold at 3.87 GeV.
Reliable anticipation of the positions of the $2P$ levels may permit their
discovery and study in the same low-energy direct-channel experiments which
studied the $1P$ levels so successfully \cite{E760}.

In order to anticipate the spin-weighted average $2P$ mass in charmonium, we
need similar quantities for the $3S$ and $1D$ levels.  The masses of the
$3^3S_1$ and $1^3D_1$ states are quoted as $4040 \pm 10$ and $3769.9 \pm 2.5$
MeV, respectively \cite{PDG}.  Assuming hyperfine splittings in the $S$-wave
levels to scale roughly as $1/N$ \cite{PL}, and using the observed splitting of
about 118 MeV between the $1^3S_1$ and $1^1S_0$ charmonium levels, we expect
the spin-weighted average $3S$ mass to be about $4030 \pm 10$ MeV. We must rely
on a specific calculation \cite{MR} of fine structure to estimate the
spin-weighted average $1D$ mass; the result is 3820 MeV.  Thus we predict the
spin-weighted $2P$ mass to lie around 3925 MeV.  The prediction of a $2P$ level
near the average of the $3S$ and $1D$ levels should be a feature of any smooth
potential which reproduces charmonium and upsilon levels.
\newpage

\leftline{\bf B.  $1D$ states of the upsilon family}
\bigskip

Electromagnetic transitions from the $\Upsilon(3S)$ to the $\chi_b'(2P)$
levels, followed by transitions to the $\Upsilon(1D)$ levels and their
subsequent decays, can give rise to characteristic photon spectra \cite{KR} to
which the CLEO-II detector is uniquely sensitive.  It is also possible for the
CLEO-II detector to scan in energy for the $\Upsilon(1^3D_1)$ level, whose
leptonic width is expected to be one or two electron volts \cite{MR}. We expect
the spin-weighted $\Upsilon(1D)$ mass to be the same distance below the
spin-weighted $2P$ mass ($10.260 \pm 0.001$ GeV) as the distance between this
mass and the spin-weighted $3S$ mass.  Here we do not know the hyperfine
splitting between the observed $\Upsilon(3S)$ at $10.3553 \pm 0.0005$ GeV and
its spin-singlet partner.  On the basis of the range of predictions for the
$1S$ level and our assumption that hyperfine splittings scale as $1/N$, we
estimate the hyperfine splitting in the $3S$ system to be 20 MeV, give or take
a factor of 2, and hence the spin-averaged $3S$ level to lie at $10.349 \pm
0.007$ GeV.  Thus, we would expect the spin-averaged $\Upsilon(1D)$ level to
lie around 10.17 GeV.

\bigskip

\leftline{\bf C.  $2D$ and $3P$ states of the upsilon family}
\bigskip

It may be possible to detect $2D$ upsilon levels through direct-channel
$e^+ e^-$ annihilation or hadronic production \cite{KR}.  Potential models
predict the spin-average of the $2D$ levels to lie around 10.44 GeV.  Let us
imagine that such a level has indeed been seen.  Then, since the $\Upsilon(4S)$
state has a mass of about 10.58 GeV, the $3P$ level should lie midway between
the two in a power-law potential, at 10.51 GeV.  Since this is below $B
\overline{B}$ threshold, the $3P$ level should be narrow, as has been noted
previously \cite{KR}.  This level, more easily produced in hadronic
interactions than the $\Upsilon(3S)$, should then be able to populate the $3S$
level through electric dipole transitions, which may be detectable.
Whichever level, $2D$ or $3P$, is seen first, our equal-spacing rule permits
the other's mass to be anticipated.
\bigskip

\leftline{\bf D.  Possible sources of error in predictions}
\bigskip

Explicit potential models predict a slightly lower $\Upsilon(1D)$ mass (ranging
from 10.15 to 10.16 GeV) than our prediction of 10.17 GeV, as a consequence of
the change in the effective power of the potential with distance.  The $3S -
2P$ splitting is more sensitive to the short-distance (more Coulomb-like) part
of the potential, while the $2P - 1D$ splitting is sensitive to longer-range
effects, where the potential is expected to be closer to linear.  For example,
a typical Coulomb-plus-linear potential, such as $V(r) = - 0.4/r + 0.16 r$,
where $r$ is in GeV$^{-1}$, and $m_b = 5$ GeV, implies $E(3S) - E(2P)$ around
15\% less than $E(2P) - E(1D)$. To a lesser degree, such changes in effective
power of the potential may be visible in deviations of the charmonium $2P$ and
upsilon $3P$ masses from our predictions.

Coupled-channel effects can significantly distort predictions of potential
models near pair production thresholds \cite{BE}.  We expect such effects to
be most significant in the anticipation of the charmonium $2P$ and
upsilon $3P$ levels mentioned above, and less important for the upsilon
$1D$ level.  \bigskip

\centerline{\bf V.  SUMMARY}
\bigskip

Interesting questions remain to be settled in quarkonium systems.  The
nonrelativistic Schr\"odinger equation continues to provide a useful guide
to the properties of such systems, with power-law potentials $V \sim r^\nu$,
$\nu \approx 0$, permitting rapid anticipation of charmonium and upsilon
properties. We have shown that in a wide range of power-law potentials,
energy levels characterized by the same principal quantum number are
approximately linear in orbital angular momentum $\ell$, with a coefficient
which is negative as long as $\nu > -1$. We have exhibited this behavior
numerically, discussed the limiting cases of perturbations around the Coulomb
and harmonic oscillator potentials, presented semiclassical results for the
coefficient of $\ell$, and applied the results to the anticipation of several
charmonium and upsilon levels which have yet to be observed.
\bigskip

\centerline{\bf ACKNOWLEDGMENTS}
\bigskip

The authors would like to thank Jeanne Rostant and Luis Alvarez-Gaum\'e for
organizing a celebration which was the starting point of this work. J. R. and
A. M. respectively wish to thank the CERN Theory Group and Rockefeller
University Physics Department for their hospitality.  This work was supported
in part by the United States Department of Energy under Contract No. DE FG02
90ER40560.
\bigskip

\newcommand{\be}{\begin{eqnarray}}
\newcommand{\en}{\end{eqnarray}}

\centerline{\bf APPENDIX A:  PROOF OF THE BAUMGARTNER THEOREM}
\centerline{\bf AND EFFECT OF LINEAR PERTURBATIONS}
\bigskip

The Schr\"odinger equation for a pure Coulomb potential is taken to be
$$
H_\ell~u_{N,\ell} = \left( -\frac{d^2}{dr^2} + \frac{\ell(\ell+1)}{r^2}
- \frac{1}{r} \right) u_{N,\ell} = - \frac{1}{4N^2} u_{N,\ell}~~~.
$$
To prove the Baumgartner theorem, we let
$$
\Delta_{N,\ell} = E_{N,\ell} - E_{N,\ell + 1}~~~.
\eqno{(A1)}
$$
Then we have
$$
\Delta_{N,\ell} = \int v \left(u^2_{N,\ell} - u^2_{N,\ell + 1}\right) dr~~~,
\eqno{(A2)}
$$
where the $u$'s are Coulomb wave functions. $u_{N,\ell}$ and $u_{N,\ell + 1}$
are linked by raising and lowering operators:
$$
A^\pm_\ell \equiv \pm  \frac{d}{dr} - \frac{\ell + 1}{r} + \frac{1}{2 (\ell +
1)}
$$
$$
A^-_\ell u_{N,\ell + 1} = \alpha_\ell u_{N,\ell}~~, ~~~A^+_\ell
u_{N,\ell} = \alpha_\ell u_{N,\ell + 1}~~~,
\eqno(A3)
$$

\medskip
\noindent
with $4 \alpha^2_\ell \equiv (\ell + 1)^{-2} - N^{-2}$.  The the Coulomb
Hamiltonian can be written as
$$
H_\ell = A_\ell^- A_\ell^+ - \frac{1}{4(\ell+1)^2}~~~.
$$
Using these operators, one can transform $\Delta_{N,\ell}$ into
$$
\alpha_\ell \Delta_{N,\ell} = \int^\infty_0 V^\prime u_{N,\ell + 1} u_{N,\ell}
dr
\eqno{(A4)}
$$
and
$$
\alpha_{\ell + 1}\Delta_{N,\ell + 1} = \int^\infty_0 V^\prime u_{N,\ell
+ 1} u_{N,\ell + 2} dr = \int^\infty_0 V^\prime u_{N,\ell + 1}
\frac{1}{\alpha_{\ell+1}}A^+_{\ell+1} u_{N,\ell+1} dr~~~.
$$
Hence, using the expression for $A^+_{\ell + 1}$ and integrating by parts,
$$
\alpha^2_{\ell + 1} \Delta_{N,\ell + 1} = \int^\infty_0 \left[-
\frac{V^{\prime\prime}}{2} - \frac{\ell + 2}{r} V^\prime + \frac{1}{2(\ell
+ 2)} V^\prime\right] u^2_{N,\ell + 1} dr~~~.
\eqno{(A5)}
$$
Similarly
$$
\alpha_\ell \Delta_{N,\ell}  = \int^\infty_0 V^\prime \left[
\frac{1}{\alpha_\ell} A^-_\ell u_{N,\ell + 1}\right] u_{N,\ell + 1}$$
and
$$
\alpha^2_\ell \Delta_{N,\ell} = \int^\infty_0 \left[ +
\frac{V^{\prime\prime}}{2} - \frac{\ell + 1}{r} V^\prime +
\frac{1}{2(\ell + 1)} V^\prime\right] u^2_{N,\ell + 1} dr~~~.
\eqno{(A6)}
$$
Combine now equations (A5) and (A6):
$$
(\ell + 1) \alpha^2_\ell \Delta_{N,\ell} - (\ell + 2) \alpha^2_{\ell +
1} \Delta_{N,\ell + 1}
$$
$$
= \int \left( (2\ell + 3) \frac{V^{\prime\prime}}{2} + (2\ell + 3)
\frac{V^\prime}{r}\right) u^2_{N,\ell + 1} dr
= \frac{2\ell +3}{2} \int \Delta V u^2_{N,\ell + 1} dr
\eqno(A7)
$$
which is what we wanted to prove.

Now we can use Eq.~(A6) to calculate $\Delta_{N,\ell}$ for the case of a
linear perturbation, $\lambda r$, to the Coulomb potential.  We get
$$
\alpha^2_{\ell} \Delta_{N,\ell} = \lambda \int \left( -\frac{\ell + 1}
{r} + \frac{1}{2(\ell+1)} \right) u^2_{N,\ell+1} dr~~~,
$$
and, using the virial theorem,
$$
\alpha^2_{\ell} \Delta_{N,\ell} = -\lambda \left[ 2 (\ell+1) |E^{\rm
Coulomb}_{N,\ell+1}| - \frac{1}{2(\ell+1)} \right]
= \frac{\lambda}{2} \frac{N^2 - (\ell+1)^2}{N^2 (\ell+1)}~~~.
\eqno(A8)
$$
Hence, from the definition of $\alpha_\ell$, $E_{N,\ell} - E_{N,\ell+1} =
2 \lambda (\ell+1)$, and, therefore, for $V = -(1/r) + \lambda r$,
$$
\lim_{\lambda \to 0} \left( \frac{E_{3S} - E_{2P}}{E_{2P} - E_{1D}} \right) =
\frac{1}{2}~~~.
\eqno(A9)
$$
\bigskip

\centerline{\bf APPENDIX B:  SOME COMMENTS ABOUT}
\centerline{\bf POWER POTENTIALS IN THE LARGE $\ell$ LIMIT}
\bigskip

i) Equation (\ref{eqn:ehatexpn}) has been obtained in Ref.~\cite{GMNP}, but in
this reference, the expansion parameter was $\left[\ell (\ell +
1)\right]^{-1/2}$. The advantage of re-expressing things in an expansion in
$(\ell + \frac{1}{2} )^{-1}$ is that for $\nu = 2$ the expansion stops and
gives the exact answer.

ii) Equation (\ref{eqn:largel}) gives only the leading term of the large $\ell$
behavior obtained by the semi-classical approximation, while Feldman,
Fulton and Devoto (FFD) give
$$
\hat{E} (n,\ell) \simeq \varepsilon(\nu) \left( \frac{|\nu|}{2}\right)^{-
\frac{\nu}{\nu + 2}}
\left( 1+\frac{\nu}{2}\right) \left[ \ell + \frac{1}{2} + (n + \frac{1}{2})
\sqrt{\nu + 2}\right]^\frac{2\nu}{\nu + 2}
$$
$$
\times \left[ 1 - \frac{(\nu-2)(\nu+1)}{24}
 \frac{(n+\frac{1}{2})^2}{ \left(\ell + \frac{1}{2} + (n+\frac{1}{2})
 \sqrt{2+\nu}\right)^2}\right]^{\frac{2\nu}{\nu + 2}}~~~.
\eqno(B1)
$$
When one expands this expression in inverse powers of $(\ell + \frac{1}{2}
)^{-1}$ and compares with Eq.~(\ref{eqn:ehatexpn}) which is the beginning of a
systematic asymptotic expansion in $(\ell + \frac{1}{2})^{-1}$ one finds a
small difference which can be corrected by subtracting 1/12 from $(n +
\frac{1}{2})^2$ in the second bracket:
$$
\hat{E} (n,\ell) \simeq \varepsilon(\nu) \left(\frac{|\nu|}{2}\right)^{
-\frac{\nu}{\nu+2}}
\left( 1+\frac{\nu}{2}\right) \left[\ell + \frac{1}{2} + (n+\frac{1}{2})
\sqrt{\nu+2}\right]^{\frac{2\nu}{\nu+2}}
$$
$$
\times \left[ 1-\frac{(\nu-2)(\nu+1)}{24} \frac{(n+\frac{1}{2})^2 -
\frac{1}{12}}{\left(\ell+\frac{1}{2} + \sqrt{2+\nu}
(n+\frac{1}{2})\right)^2}\right]^{\frac{2\nu}{\nu+2}}~~~.
\eqno(B2)
$$
It turns out that the small change made in the second bracket makes this new
expression extremely accurate especially for small $n$ and in particular for
$n=0$.  Then for $\nu = -0.5$, $\nu = 0.5$, $\nu = 1$ we find by numerical
tests that the relative error of Eq.~(B2) is less than $10^{-4}$ for $n=0$,
arbitrary $\ell\ge 0$.  For $n=1$ it is less than $5\times 10^{-3}$.  We
believe that this holds on the whole interval $-1 \le\nu\le 2$.  Examples
of the accuracy of Eq.~(B2) for $\nu = -0.5$ and $\nu = 1$ are shown in Tables
1 and 2.

\begin{table}
\caption{Comparison of exact and approximate [Eq.~(B2)] energy levels for the
Schr{\"o}dinger equation (\protect{\ref{eqn:schr}}) in a potential $V(r) = -
r^{-1/2}$.}
\begin{center}
\begin{tabular}{r c c c c} \hline
Exact &   $N = 1$  &   $N = 2$  &   $N = 3$  &   $N = 4$  \\ \hline
S     & --0.438041 & --0.263203 & --0.197558 & --0.161705 \\
P     &            & --0.286611 & --0.209800 & --0.169416 \\
D     &            &            & --0.221506 & --0.176817 \\
F     &            &            &            & --0.184005 \\ \hline
      &            &            &            &            \\
Eq.~(B2) & $N = 1$ &   $N = 2$  &   $N = 3$  &   $N = 4$  \\ \hline
S     & --0.438043 & --0.264647 & --0.199228 & --0.163352 \\
P     &            & --0.286615 & --0.210156 & --0.170025 \\
D     &            &            & --0.221507 & --0.176947 \\
F     &            &            &            & --0.184006 \\ \hline
\end{tabular}
\end{center}
\end{table}

\begin{table}
\caption{Comparison of exact and approximate [Eq.~(B2)] energy levels for
the Schr{\"o}dinger equation (\protect{\ref{eqn:schr}}) in a potential $V(r) =
r$.}
\begin{center}
\begin{tabular}{r c c c c} \hline
Exact &  $N = 1$ &  $N = 2$ &  $N = 3$ &  $N = 4$ \\ \hline
S     & 2.338107 & 4.087949 & 5.520560 & 6.786708 \\
P     &          & 3.361255 & 4.884452 & 6.207623 \\
D     &          &          & 4.248182 & 5.629708 \\
F     &          &          &          & 5.050926 \\ \hline
      &          &          &          &          \\
Eq.~(B2) & $N=1$ &   $N=2$  &   $N=3$  &   $N=4$  \\ \hline
S     & 2.338231 & 4.066542 & 5.479175 & 6.727770 \\
P     &          & 3.361231 & 4.874358 & 6.183454 \\
D     &          &          & 4.248160 & 5.623973 \\
F     &          &          &          & 5.050910 \\ \hline
\end{tabular}
\end{center}
\end{table}

iii) The smoothness of $\hat{E} (n,\ell)$ for fixed $n$, as a function of
$\ell$ can be connected to the fact that $\left[
\hat{E}(n,\ell)\right]^{(\nu+2)/\nu}$ is a ``Herglotz'' function of $(\ell
+ \frac{1}{2})^2$.  A Herglotz function is defined by
$$
H(z) = A + Bz + \frac{z}{\pi} \int^0_{-\infty} \frac{{\rm Im} H(z^\prime)
dz^\prime}{z^\prime (z^\prime - z)}
$$
with ${\rm Im} H(z^\prime)\ge 0$. It has the characteristic property ${\rm Im}
H(z)/{\rm Im} z > 0$, and cannot grow faster than $z$ in any complex direction.
Furthermore, it is easy to see that it is concave, i.e. $d^2 H(z)/dz^2 \le 0$.

It has been shown by one of us (A. M.) and Harald Grosse [20] that
$\hat{E}(n,\ell)$ for $V=r^\nu,~ \nu>0$ is analytic in the variable
$z=(\ell + \frac{1}{2})^2$ in a cut plane, with the cut running from
$z=-\infty$ to $z=0$.  For positive $z$, $E$ is real.  Furthermore for
${\rm Im} z>0$ we have $0 < {\rm Arg}~E < \frac{\nu}{\nu+2} \pi$, hence
$0< {\rm Arg}~(E)^{(\nu+2)/\nu} <
\pi$.  Hence $E^{(\nu+2)/\nu}$ is a Herglotz function of $z$, and, as we
said, is, in particular, concave.  Let us illustrate the usefulness of
this remark by considering $V=r^4$.  We have, for instance
$$
\hat{E} (4,0) = 44.0 ~~~\hat{E}(4,1) = 50.1 ~~~\hat{E} (4,2)=56.4~~~.
$$
Using the concavity of $E^{3/2}$ in $(\ell + \frac{1}{2})^2$ we get from
$\hat{E}(4,0)$ and $\hat{E}(4,2)$ the bound
$$
\hat{E} (4,1) > 48.3~~~.
$$
\bigskip

\centerline{\bf APPENDIX C:}
\centerline{\bf BEHAVIOR OF $R_{N,\ell}$ NEAR $\nu = -1$ AND $\nu=2$}
\bigskip

To study the behavior of $R_{N,\ell}$ in the neighborhood of $\nu=-1$ we
have to take a perturbation of the Coulomb potential which is $v =
(\log r)/r$, but we shall first look at a perturbing potential which is
$v = \log r$.  Then according to equation (A6) of Appendix A we
have
$$
\alpha^2_\ell \Delta_{N,\ell} = \left\langle \frac12 v^{\prime\prime} -
\frac{\ell +
1}{r} v^\prime + \frac{1}{2 (\ell + 1)} v^\prime \right\rangle_{N,\ell+1}
= - \frac{2\ell + 3}{2} \left\langle\frac{1}{r^2}\right\rangle_{N,\ell+1} +
\frac{1}{2(\ell + 1)} \left\langle\frac{1}{r}\right\rangle_{N,\ell+1}
$$
$$
= - \frac12 \frac{\partial E}{\partial\ell} |_{\rm n~ fixed} +
\frac{1}{2(\ell+1)} (-2E)
= \frac{1}{4N^2} \left( \frac{1}{\ell + 1} - \frac{1}{N}\right)~~~.
\eqno(C1)
$$
Therefore, for $v=\log r$,
$$
\Delta_{N,\ell} = \frac{1}{N} \frac{(\ell+1)}{(N+\ell+1)} =
\int^\infty_0 \left( u^2_{N,\ell} - u^2_{N,\ell+1}\right) \log r dr~~~.
\eqno(C2)
$$

Now, following a standard strategy of multiplying the Schr\"odinger
equation by $u \log r$ and $u^\prime r\log r$ and integrating, it is
possible to obtain the following identity:
$$
\int^\infty_0 \left( u^2_{N,\ell} - u^2_{N,\ell+1}\right) \frac{\log r}
{r} dr = \frac{1}{2N^2} \int^\infty_0 \left(u^2_{N,\ell} -
u^2_{N,\ell+1}\right) \log r dr - \frac{1}{2N^3}~~~.
\eqno(C3)
$$
Hence combining (C2) and (C3) we get
$$\int^\infty_0 \left(u^2_{N,\ell} - u^2_{N,\ell+1}\right) \frac{\log r}
{r} dr = - \frac{1}{2N^2 (N+\ell+1)}
\eqno(C4)
$$
and hence
$$
R_{N,\ell} = \frac{E_{N,\ell-1} - E_{N,\ell}}{E_{N,\ell} - E_{N,\ell+1}}
= \frac{N+\ell+1}{N+\ell}~~~.
\eqno(C5)
$$

Now we turn to the neighborhood of $\nu = 2$.  The harmonic oscillator
Hamiltonian is taken to be
$$h_\ell = - \frac{d^2}{dr^2} + \frac{\ell (\ell + 1)}{r^2} + \frac{r^2}{4}
{}~~.
\eqno(C6)
$$
Then, for $V=r^2/4$, we have $\hat{E} (n,\ell) = 2n + \ell + \frac{3}{2}$, and
$\hat{E} (n+1,\ell - 1) - \hat{E} (n,\ell) = 1$.  We consider now the
perturbation
$$
\delta_\ell = \int^\infty_0 v \left(u^2_{n,\ell} - u^2_{n-1,
\ell+1}\right) dr~~~.
\eqno(C7)
$$
With
$$
B^\pm_\ell \equiv \pm \frac{d}{dr} - \frac{\ell + 1}{r} + \frac{r}{2}
\eqno(C8)
$$
we have
$$
B^+_\ell u_{n+1,\ell} = \gamma_n u_{n, \ell + 1} ~~~,
B^-_\ell u_{n,\ell+1} = \gamma_n, u_{n+1,\ell}~~~,
\eqno(C9)
$$
with $\gamma^2_n \equiv 2n+2$.
Then, by repeated use of these raising and lowering operators one can get
$$
\gamma^2_{n-1} \delta_\ell = \int^\infty_0 \frac{v^\prime}{r} \left[
\left(\frac{ru^2}{2}\right)^\prime - \frac{u^2}{2} - (\ell+1)
u^2+\frac{r^2}{2} u^2\right] dr~~~,
$$
$$\gamma^2_n \delta_{\ell - 1} = \int^\infty_0 \frac{v^\prime}{r} \left[
\left( -\frac{ru^2}{2}\right)^\prime + \frac{u^2}{2} - \ell u^2 +
\frac{r^2}{2} u^2\right] dr~~~,
$$
where $u \equiv u_{n,\ell}$. Then we get
$$
\delta_\ell - \delta_{\ell-1} = \int^\infty_0 -
\left(\frac{v^\prime}{r}\right)^\prime \frac{2n+1}{4n(n+1)}
r u^2 dr
$$
$$
+ \frac{1}{2n(n+1)} \int^\infty_0 \frac{v^\prime}{r}
\left[\frac{r^2}{2} - E(n,\ell)\right] u^2 dr~~~.
\eqno(C10)
$$

For $v=\log r$ and for $v=r^2\log r$ all these integrals can be calculated and
one finds in both cases $\delta_\ell - \delta_{\ell - 1} = 0$. Hence if the
total potential is
$$
V = r^\nu = r^2 + r^\nu - r^2 \simeq r^2 + (\nu - 2) r^2\log r~~~,
$$
then $\delta_\ell - \delta_{\ell - 1}$ is of higher order, i.e. ${\cal O}
(\nu - 2)^2$, which fits with the numerical observations of Fig.~3.
\bigskip

\def \ajp#1#2#3{Am. J. Phys. {\bf#1}, #2 (#3)}
\def \apny#1#2#3{Ann. Phys. (N.Y.) {\bf#1}, #2 (#3)}
\def \app#1#2#3{Acta Phys. Polonica {\bf#1}, #2 (#3)}
\def \arnps#1#2#3{Ann. Rev. Nucl. Part. Sci. {\bf#1}, #2 (#3)}
\def \cmts#1#2#3{Comments on Nucl. Part. Phys. {\bf#1}, #2 (#3)}
\def \cn{Collaboration}
\def \cp89{{\it CP Violation,} edited by C. Jarlskog (World Scientific,
Singapore, 1989)}
\def \efi{Enrico Fermi Institute Report No. EFI}
\def \f79{{\it Proceedings of the 1979 International Symposium on Lepton and
Photon Interactions at High Energies,} Fermilab, August 23-29, 1979, ed. by
T. B. W. Kirk and H. D. I. Abarbanel (Fermi National Accelerator Laboratory,
Batavia, IL, 1979}
\def \hb87{{\it Proceeding of the 1987 International Symposium on Lepton and
Photon Interactions at High Energies,} Hamburg, 1987, ed. by W. Bartel
and R. R\"uckl (Nucl. Phys. B, Proc. Suppl., vol. 3) (North-Holland,
Amsterdam, 1988)}
\def \ib{{\it ibid.}~}
\def \ibj#1#2#3{~{\bf#1}, #2 (#3)}
\def \ichep72{{\it Proceedings of the XVI International Conference on High
Energy Physics}, Chicago and Batavia, Illinois, Sept. 6 -- 13, 1972,
edited by J. D. Jackson, A. Roberts, and R. Donaldson (Fermilab, Batavia,
IL, 1972)}
\def \ijmpa#1#2#3{Int. J. Mod. Phys. A {\bf#1}, #2 (#3)}
\def \ite{{\it et al.}}
\def \lkl87{{\it Selected Topics in Electroweak Interactions} (Proceedings of
the Second Lake Louise Institute on New Frontiers in Particle Physics, 15 --
21 February, 1987), edited by J. M. Cameron \ite~(World Scientific, Singapore,
1987)}
\def \ky85{{\it Proceedings of the International Symposium on Lepton and
Photon Interactions at High Energy,} Kyoto, Aug.~19-24, 1985, edited by M.
Konuma and K. Takahashi (Kyoto Univ., Kyoto, 1985)}
\def \mpla#1#2#3{Mod. Phys. Lett. A {\bf#1}, #2 (#3)}
\def \nc#1#2#3{Nuovo Cim. {\bf#1}, #2 (#3)}
\def \np#1#2#3{Nucl. Phys. {\bf#1}, #2 (#3)}
\def \pisma#1#2#3#4{Pis'ma Zh. Eksp. Teor. Fiz. {\bf#1}, #2 (#3) [JETP Lett.
{\bf#1}, #4 (#3)]}
\def \pl#1#2#3{Phys. Lett. {\bf#1}, #2 (#3)}
\def \plb#1#2#3{Phys. Lett. B {\bf#1}, #2 (#3)}
\def \pr#1#2#3{Phys. Rev. {\bf#1}, #2 (#3)}
\def \prd#1#2#3{Phys. Rev. D {\bf#1}, #2 (#3)}
\def \prl#1#2#3{Phys. Rev. Lett. {\bf#1}, #2 (#3)}
\def \prp#1#2#3{Phys. Rep. {\bf#1}, #2 (#3)}
\def \ptp#1#2#3{Prog. Theor. Phys. {\bf#1}, #2 (#3)}
\def \rmp#1#2#3{Rev. Mod. Phys. {\bf#1}, #2 (#3)}
\def \rp#1{~~~~~\ldots\ldots{\rm rp~}{#1}~~~~~}
\def \si90{25th International Conference on High Energy Physics, Singapore,
Aug. 2-8, 1990}
\def \slc87{{\it Proceedings of the Salt Lake City Meeting} (Division of
Particles and Fields, American Physical Society, Salt Lake City, Utah, 1987),
ed. by C. DeTar and J. S. Ball (World Scientific, Singapore, 1987)}
\def \slac89{{\it Proceedings of the XIVth International Symposium on
Lepton and Photon Interactions,} Stanford, California, 1989, edited by M.
Riordan (World Scientific, Singapore, 1990)}
\def \smass82{{\it Proceedings of the 1982 DPF Summer Study on Elementary
Particle Physics and Future Facilities}, Snowmass, Colorado, edited by R.
Donaldson, R. Gustafson, and F. Paige (World Scientific, Singapore, 1982)}
\def \smass90{{\it Research Directions for the Decade} (Proceedings of the
1990 Summer Study on High Energy Physics, June 25--July 13, Snowmass,
Colorado),
edited by E. L. Berger (World Scientific, Singapore, 1992)}
\def \tasi90{{\it Testing the Standard Model} (Proceedings of the 1990
Theoretical Advanced Study Institute in Elementary Particle Physics, Boulder,
Colorado, 3--27 June, 1990), edited by M. Cveti\v{c} and P. Langacker
(World Scientific, Singapore, 1991)}
\def \yaf#1#2#3#4{Yad. Fiz. {\bf#1}, #2 (#3) [Sov. J. Nucl. Phys. {\bf #1},
#4 (#3)]}
\def \zhetf#1#2#3#4#5#6{Zh. Eksp. Teor. Fiz. {\bf #1}, #2 (#3) [Sov. Phys. -
JETP {\bf #4}, #5 (#6)]}
\def \zpc#1#2#3{Zeit. Phys. C {\bf#1}, #2 (#3)}

\end{document}